\documentclass[11pt]{article}
\usepackage[left=2.5cm,top=2.5cm,right=2cm,bottom=2.5cm]{geometry}
\usepackage{amsmath, amssymb}

\usepackage{graphicx,subfigure,epsfig,epsf,color}
\usepackage[utf8x]{inputenc}
\usepackage{slashed}
\usepackage{graphicx}
\usepackage{graphics}
\usepackage{epstopdf}
\usepackage{subfigure}
\usepackage{amsfonts}
\usepackage{sectsty}
\usepackage{hyperref}
\usepackage{cite}
\usepackage{pdflscape}
\usepackage{lipsum}
\newcommand{\cmmnt}[1]{\ignorespaces}

\def\dnul{\partial_{\nu}}

\newcommand{\g}{\gamma}
\newcommand{\om}{\omega}

\newcommand{\beq}{\begin{equation}}
\newcommand{\eeq}{\end{equation}}
\newcommand{\bea}{\begin{eqnarray}}
\newcommand{\eea}{\end{eqnarray}}
\newcommand{\beas}{\begin{eqnarray*}}
\newcommand{\eeas}{\end{eqnarray*}}
\newcommand{\bcr}{\begin{center}}
\newcommand{\ecr}{\end{center}}
\newcommand{\vphi}{\langle \phi \rangle}
\newcommand{\etal}{\textit{et. al.,}}
\begin{document}
\date{}

\title{SN1987A cooling due to Plasmon-Plasmon scattering in the Randall-Sundrum Model}
\maketitle
\begin{center}
\author{Manish Kumar Sharma}\footnote{p20190006@goa.bits-pilani.ac.in},~Saumyen Kundu\footnote{p20170022@goa.bits-pilani.ac.in}, Prasanta Kumar Das\footnote{pdas@goa.bits-pilani.ac.in} \\
 \end{center}
 
 \begin{center}
 Department of Physics \\
 Birla Institute of Technology and Science-Pilani, K. K. Birla Goa campus, \\
 NH-17B, Zuarinagar, Goa-403726, India
 \end{center}
 \vspace*{0.25in}
 

%
\begin{abstract} 
The light braneworld radion, stabilized via the Goldberger-Wise mechanism in the Randall-Sundrum model, can be produced copiously inside the supernova core due to plasmon-plasmon annihilations. The radion, thus produced, subsequently decays to  a neutrino-antineutrino pair and takes away the energy released in the SN1987A explosion.  Assuming that the supernovae cooling rate for any new physics channel(Raffelt's criterion) $\dot{\varepsilon}(\gamma_P\gamma_P  \stackrel{\phi}{\longrightarrow}  \nu_{\tau} \overline{\nu}_{\tau}) \leq 7.288\times 10^{-27} \rm{GeV}$, we find the lower bound on the radion vacuum expectation value $\vphi \sim 3$~TeV for $m_\phi = 30$ MeV corresponding to the deformation parameter $q=1.17$ in the Tsallis statistics formalism. In the scenario with $q=1$, we find $\vphi = 178$ GeV for $m_\phi = 30$ MeV.\\

\end{abstract}
%
\section{Introduction }

Numerous theoretical models have been proposed to explain the large gap between the Planck scale $M_{Pl}\;(\sim 10^{19}~ {\rm GeV}$) and the electro-weak scale $M_{EW}(\sim 100~ {\rm GeV})$ by assuming extra spatial dimensions of factorizable as well as non-factorizable geometry~\cite{ADD, Giudice:2000av, RS, Goldberger:1999un, Goldberger:2000dv, Datta:2013lja, Kouwn:2017qet, OliveiradosSantos:2019rjt, Dasgupta:2018zgi, Kawai:2019uei, Khlopunov:2022jaw, Tan:2022vfe, Matsumoto:2022fln, Kakizaki:2021kof, Li:2021pnv, Borici:2020lpu, Mathieu:2020ywc}. 
\cmmnt{To explain the large gap between the Planck scale $M_{Pl}\;(\sim 10^{19}~ {\rm GeV}$) and the electro-weak scale $M_{EW}(\sim 100~ {\rm GeV})$, a plethora of models based on extra spatial dimensions {\bf of factorizable and non-factorizable geometry}, have been proposed~\cite{ADD, Giudice:2000av, RS, Goldberger:1999un, Goldberger:2000dv, Datta:2013lja, Kouwn:2017qet, OliveiradosSantos:2019rjt, Dasgupta:2018zgi, Kawai:2019uei, Khlopunov:2022jaw, Tan:2022vfe, Matsumoto:2022fln, Kakizaki:2021kof, Li:2021pnv, Borici:2020lpu, Mathieu:2020ywc}. }
Randall-Sundrum (RS) model among them is quite interesting as it solves  the hierarchy problem quite elegantly \cite{RS}. The model considers the world in a $5$-dimensional space where the fifth spatial dimension is characterized by the angular coordinate $- \pi \le \theta \le \pi$ and the space is $S^1/Z_2$ orbifold (i.e. the point $(x,\theta)$ is identified with the point $(x,-\theta)$). The metric describing such a $5$-dimensional world is non-factorizable and a line element in this space-time is given by
\bea \label{eqn:rsmetric}
d s^2 = e^{-2 k R_c |\theta|}\eta_{\mu \nu} d x^\mu d x^\nu
- R_c^2 d \theta^2
\eea
\noindent
where $k$ is the bulk curvature constant and $x^\mu$ are the Lorentz coordinates of four dimensional surfaces of constant $\theta$. This theory postulates two $D_3$ branes to be  placed along $x^\mu$ directions \cmmnt{living} in a $5$-dimensional world: one is located at the orbifold point $\theta = 0$ where gravity peaks and the other one is at the orbifold point $\theta = \pi$ where the Standard Model (SM) fields exist and gravity is weak. The factor $e^{- 2 k R_c |\theta|}$ appearing in the metric in Eq. \eqref{eqn:rsmetric} is known as the warp factor. 
 The compactification radius $R_c$($\sim$ the distance between the two $D_3$ branes) can be related to the vacuum expectation value (VEV) of the modulus field $T(x)$ which corresponds to the fluctuations of the metric over the background geometry given by $R_c$. Replacing $R_c$ by the modulus field $T(x)$, we can rewrite 
the RS metric at the orbifold point $\theta = \pi$ as

\bea
d s^2 = g_{\mu \nu}^{vis} d x^\mu d x^\nu - T(x)^2 d \theta^2
\eea
where, $g_{\mu \nu}^{vis} = e^{- 2 \pi k T(x)}\eta_{\mu \nu}$.
We write $e^{- 2 \pi k T(x)}= \left(\frac{\Phi(x)}{f}\right)^2 \eta_{\mu \nu}$, where $f^2 = \frac{24 M_5^3}{K}$, $M_5$ is the $5$-dimensional Planck scale and $\Phi(x) = \langle \phi \rangle + \phi(x)$, where $\phi(x)$ is dubbed as the radion fluctuation and $\langle \phi \rangle$ is the vacuum expectation value (VEV) of the modulus field $T(x)$. The stabilization of the modulus field is done by a potential using a bulk scalar with suitable interactions with the two $3$ branes located at two orbifold points at $\theta = 0$ and $\theta = \pi$,  which is called the Goldberger-Wise mechanism \cite{GW}, and one ends up with a radion of nonzero mass \cite{GW, PhysRevD.60.107505, Giudice:2000av,Goldberger:1999un, Goldberger:2000dv}. 
\par 
The radion can be lighter than the other low-lying gravitonic degrees of freedom and can emerge as the first messenger of the compact extra spatial dimension \cmmnt{in collider experiments}. Several studies on the observable implications of radion in collider and astrophysical contexts are available in the literature. 
\cite{Graesser,Mahanta:2000zp,Csaki:1999mp,Mahanta:457152,Park:2001uc,Kim:2001rc,Cheung:2000rw,Chaichian:2001rq,Das:2001pn,Das:2005na}.
Among the various astrophysical phenomenon where the lack of clear understanding requires the \cmmnt{intervention} introduction of new physics, i.e., physics beyond the Standard Model of Particle Physics, the core-collapse supernova SN1987A deserves special attention.
\par
The core-collapse supernovae are the class of explosions that mark the evolutionary end of massive stars ($M \ge 8\, M_\odot$) and SN1987A is an example of a core-collapse supernova. The energy liberated in a supernova explosion is the gravitational binding energy and it is about $3\times10^{53}~{\rm ergs}$, 99\% of which goes into neutrinos and the remaining 1\% to the kinetic energy of the explosion. The detection of these neutrinos by the earth-based detector is the main astroparticle interest \cmmnt{of the} {\bf in the context of} core-collapse supernovae.

The neutrino flux emitted by the supernova SN1987A explosion  was measured by Kamiokande \cite{Kamiokande-II:1987idp} and IMB \cite{Bionta:1987qt} collaborations. The measured energies which follow from the SN1987A data are found to be ``too low" in comparison to the values obtained from the numerical neutrino light curves. This raises the possibility of whether the anomaly found leads to a serious problem with the SN models or the detectors, or is their new physics happening in supernovae \cite{Das:2001pn, Das:2005na} 
In the case of SN1987A, the entire gravitational binding energy of about $10^{53}~{\rm ergs}$ was released just in a few seconds, and the neutrino fluxes were measured by Kamiokande \cite{Kamiokande-II:1987idp} and IMB \cite{Bionta:1987qt} collaborations. The measured energies which follow from the SN1987A data are found to be ``too low" in comparison to the values obtained from the numerical neutrino light curves. For example, the numerical simulation in \cite{Totani:1997vj} yields time-integrated values $\langle E_{\nu_e}\rangle\approx13~{\rm MeV}$, $\langle E_{\bar\nu_e}\rangle\approx16~{\rm MeV}$, and $\langle E_{\nu_x}\rangle\approx23~{\rm MeV}$.  On the other hand, the data imply $\langle E_{\bar\nu_e}\rangle=7.5~{\rm MeV}$ at Kamiokande and 11.1~MeV at IMB~\cite{Jegerlehner:1996kx}.  Even the 95\% confidence range for Kamiokande implies $\langle E_{\bar\nu_e}\rangle<12~{\rm MeV}$.  
Flavor oscillations would increase the expected energies and thus enhance the discrepancy~\cite{Jegerlehner:1996kx}.  
The question is whether It has remained unclear if these and other anomalies of the SN1987A neutrino signal should be blamed on small-number statistics, or point to a serious problem with the SN models or the detectors, or if is there new physics happening in supernovae.
\par
The role of new physics to understand the SN1987A energy loss dynamics have drawn a lot of attention among the physics community (see \cite{Raffelt}). In particular, the impact of extra spatial dimension(s) on the supernova energy loss rate has been investigated by several groups (see Das \etal \cite{Das:2001pn, Das:2005na} and references therein).   
In this work we investigate one such neutrino production mechanism where the neutrino pairs are produced by a light stabilized radion in the Randall-Sundrum model\cite{RS, GW}. The radion, produced in the outer crust (comparatively less dense than the inner one) of the supernova core due to plasmon-plasmon annihilation \footnote{Plasmons, in contrast to the ordinary photons which are transverse, are the quanta of the electromagnetic field in plasma and have two transverse polarization vectors and one longitudinal polarization vector perpendicular to the wave vector subsequently decays to a neutrino pair which freely escapes the star and takes away the energy released in the explosion\cite{Leinson, Kantor}. We work within the framework of Tsallis statistics\cite{Tsallis} to take into account the fluctuation of the supernova temperature.} 

The outline of this work is as follows. In Sec.~\ref{sec:IIA}, we briefly discuss the Supernova SN1987A cooling and give a short review of the existing work on supernova cooling and new physics. We give a small introduction to Tsallis statistics in Sec.~\ref{sec:IIB} 
In Sec.~\ref{sec:III}, we discuss the role of a light stabilized radion in SN1987A cooling due to neutrino pair production in the Randall Sundrum model within the framework of Tsallis statistics formalism with $q \neq 1$(deformed scenario) and $q = 1$ (undeformed scenario). The numerical analysis is presented in Sec.~\ref{sec:IV}. Using Raffelt's criterion, we obtain a lower bound on the radion vev, $\vphi$, corresponding to different radion mass, $m_\phi$. Finally, in Sec.~\ref{sec:V}, we summarize our results and conclude.  
\section{Supernova Explosion, its cooling and Tsallis statistics}
\subsection{Supernova Cooling, Raffelt's criterion} \label{sec:IIA}

If we propose some novel new physics channel through which the supernova SN1987A can lose energy, the luminosity of that channel should be such (low) that it agrees with the neutrino observations with theory i.e. the channel luminosity $${\cal L}_{new\, channel} \le 10^{53}\, ergs\, s^{-1}$$ 
Based on a detailed supernova simulation, Raffelt~\cite{Raffelt} has proposed a simple analytic criterion: if the emissivity of any energy-loss mechanism is greater than $10^{19}$ ergs g$^{-1}$ s$^{-1}$ then it will remove sufficient energy from the explosion to invalidate the current understanding of Type-II supernovae's neutrino signal. 

\noindent This constraint on the emissivity of a given process can be converted into a bound on the new physics parameters - in case of large extra dimension, a  bound on the D($=4+d$) dimensional Planck scale $M_D$ (where $d$, the number of extra spatial dimensions) (ADD model\cite{ADD}) or the radion vev $\langle \phi \rangle$ of warped geometry (Randall-Sundrum model\cite{RS}).
 Considering the temperature of the outer crust of the SN1987A core to be $T=30$ MeV  and its density $\rho_{SN}=3 \times 10^{14}$ g cm$^{-3}$~\footnote{Note that the temperature of the innermost core of SN1987A is $T = 70$ MeV, while its innermost core density is $\rho_{SN} = 10 \times 10^{14}$  g cm$^{-3}$}, we have shown in Table \ref{table:1}, the bound on $M_D$ obtained by various authors. 
\begin{table}[t]
\centering
\begin{tabular}{|l|c|c|}
\hline
Group/Collaboration & $M_D$ (GeV) & $d$   \\
\hline\hline
Cullen \etal \cite{Cullen:1999hc} & $\ge 50$ TeV,~ $\ge 4$ TeV,~$\ge 1$ TeV & 2,~3,~4 \\
Barger \etal \cite{Barger:1999jf} & $\ge 51$ TeV,~ $\ge 3.6$ TeV & 2,~3 \\
Hannestad \etal \cite{Hannestad:2001jv} & $\ge 84$ TeV,~$\ge 7$ TeV & 2,~3 \\
\hline
\end{tabular}
\caption {\label{table:1}The lower bound on $M_D$ for different $d$ is shown in this table, which follows from Raffelt's criteria on the supernova SN1987A energy loss rate for any new channel i.e. $\dot{\varepsilon} \le 7.288\times 10^{-27} \rm{GeV}$.}
\end{table}

\noindent In addition, the KK gravitons produced in plasmon-plasmon collision inside the supernovae may also contribute to its cooling, and this gives rise to a strong bound on $M_D$, e.g., for $d=2$ one finds $M_D \ge 22.5~{\rm TeV}$, while for $d=3$, it is $M_D \ge 1.4~{\rm TeV}$ \cite{Das:2008ss}. \\
\noindent Randall-Sundrum (RS) model, the second variant of the extra-dimensional models,  can also play a crucial role in supernova physics. Uma \etal \cite{Mahanta:2000mx} studied the impact of a light radion on neutrino-antineutrino oscillation. They found that for a light radion of mass $m_\phi \ge 1~{ {{\rm GeV}}}$ with $\vphi = 1~{\rm{TeV}}$, the interaction potential(arising due to the exchange of a radion between the supernova matter and the neutrino-antineutrino pair) does not affect the neutrino oscillation.  Although quite a few works on new physics in the context of SN1987A energy loss rate are available in the literature, a study of this kind with a light stabilized radion in the presence of fluctuating core temperature of the SN1987A is still lacking. The present work in the braneworld scenario with a stabilized radion, is an effort in that direction.\\ 
Working within the framework of Tsallis statistics \footnote{See the next section for a short review on Tsallis statistics, we investigate here how the neutrinos are produced from the decay of light and stabilized radion and then  take away the energy released in the supernova SN1987A explosion.} 
 The processes of our interest are   
 
\[\gamma_P + \gamma_P \to \phi \to \nu_\ell + \overline{\nu}_\ell\] 

\noindent where $\ell = e, \mu, \tau$. 
 Note that the coupling of radion with other particles is proportional to their mass.  Since  $m_{\nu_e} \ll m_{\nu_\tau} $ and $m_{\nu_\mu} \ll m_{\nu_\tau}$~\cite{Belesev1994,PhysRevD.53.6065,BUSKULIC1995585}, the radion coupling to electron-neutrino ($\nu_e$) and muon-neutrino ($\nu_\mu$) are extremely small in comparison to tau-neutrino ($\nu_\tau$) and hence will not lead to a practically significant bound on $\vphi$. However, in this work, we have studied all three channels which comprise neutrinos of $e$-type, $\mu$-type, and $\tau$-type.  
\subsection{Temperature fluctuation and Tsallis statistics}\label{sec:IIB}
 The core temperature of the supernova is fluctuating and this temperature fluctuations  in the Tsallis  statistics \cite{Tsallis, Beck_cohen} takes the following  $\chi^2$ distribution form   
\bea
f(\beta) = \frac{1}{\Gamma\left(\frac{n}{2}\right)} \left(\frac{n}{2 \beta_0} \right)^{n/2} \beta^{\frac{n}{2} - 1} ~exp\left(- \frac{n \beta}{2 \beta_0}\right)
\label{fbeta}
\eea 
where $n$ is the degree of the distribution and the inverse temperature $\beta = {1 \over {k T}}$. 
The average of the fluctuating inverse temperature $\beta$ can be estimated as 
\bea
\langle \beta \rangle = n \langle X_i^2 \rangle = \int_{0}^{\infty} \beta f(\beta) d\beta = \beta_0
\eea 
Taking into account the local temperature fluctuations, integrating over all $\beta$, we find
the $q$-generalized relativistic Maxwell-Boltzmann distribution 
\bea
{\mathcal{P}}(E) \sim  \frac{E^2}{\left(1 + b(q-1)E\right)^{\frac{1}{q-1}}}
\eea
where  $q = 1 + \frac{2}{n + 6}$ (the deformation parameter) and $b = \frac{\beta_0}{4 - 3 q}$. Here $\beta_0 = \frac{1}{k T_0}$ where $T_0$ is the average temperature. It is a generalization of Fermi-Dirac 
and Bose-Einstein distribution is worked out in \cite{Beck_super}.  The average occupation number of any particle within the Tsallis statistics \cite{Tsallis}), 
is given by  $f_i(\beta,E_i)$ ($i = 1,~2$ corresponds to particles) where
\bea
f_i(\beta,E_i) = \frac{1}{\left(1 + (q-1)b E_i \right)^{\frac{1}{q-1}} \pm 1}
\eea
where the $``-"$ sign is for bosons and $``+"$ sign is for fermions. 
Note that the effective Boltzmann factor 
$x_i = \left(1 + (q-1)b E_i \right)^{-\frac{1}{q-1}}$ approaches to the ordinary Boltzmann factor 
$e^{- b E_i} (= e^{- \beta_0 E_i})$ as $q \to 1$.  
A study by C. Beck \cite{C.Beck} has shown that Tsallis distributions are observed for many systems because of the fluctuations in an intensive parameter like inverse temperature, friction constant, etc and for $\chi^2$ distributed intensive parameter, the deformation parameter $q$ has to be $q \ge 1$, whereas the range $0 < q <1$ is important in nonextensive statistical mechanics. An upper limit on $q$ has been found by comparing (anti)neutrino emission spectra from recent supernova simulations with the Tsallis spectra. The maximum value of $q$ ($q_{max}$=1.27) can be considered as the limiting value of the deformation parameter in order to fit with the recent state-of-the-art supernovae modeling \cite{Guha, Nikrant:2017, Tamborra}. The Tsallis statistics finds important applications in collider physics and astrophysics: 
Beck \etal \cite{Beck_cosmic,C.Beck} uses the $q$-deformed statistics in order to explain the 
measured energy spectrum of primary cosmic rays. With $b^{-1} = k T_0 = 107~\rm{MeV}$ and the 
deformation parameter $q = 1.215$  and $q = 1.222$, respectively, they were able to explain quite well the flux rate, i.e., the upper (upto the knee) portion and the lower (the ankle) portion of the cosmic ray spectrum \cite{Beck_super}. In Ref.~\cite{Bediaga-ep}, the authors used the $q$-deformed statistics to explain the differential cross-section for transverse momenta in electron-positron annihilation.  The applications of $q$-deformed statistics for chaotically  quantized scalar fields \cite{Beck_deform}, dark energy \cite{Beck_dark} are available in the literature.
In a work, Das \etal  \cite{Das} found that an ultra-light radion can explain the SN1987A energy loss rate provided $q$ lies within the range $1.18 < q < 1.32$.  
\section{SN1987A cooling : Neutrino pair production in Randall-Sundrum model} \label{sec:III}
Since the radion coupling to the matter is determined by the $4-d$ general covariance, it couples to  the trace of the energy-momentum tensor of the matter (standard model) fields which resides on the TeV brane \cite{RS, GW}
\beq
{\cal L}\ = \frac{\phi}{\vphi} T^{\mu}_{\mu}\
\label{interaction}
\eeq
where the radion vev $\vphi \sim {\rm TeV}$ for $k R_c \simeq 12$ (with $k$ and $R_c$ as the bulk curvature constant and the size of the fifth spatial dimension, respectively) in order to \cmmnt{produce} obtain the weak scale from the Planck scale through the exponential warp factor, $e^{-\pi k R_c}$ \cite{RS, GW} and $T^{\mu}_{\mu}$ of the matter fields (which includes SM fermions, gauge fields, and Higgs boson, etc) which can be written as  
\bea
T^\mu_\mu  = \sum_{\psi} \left[\frac{3 i}{2} \left({\overline{\psi}}
\g_\mu \dnul \psi - \dnul{\overline{\psi}} \g_\mu \psi \right)\eta^{\mu\nu}
- 4 m_\psi {\overline{\psi}} \psi\right] - 2 m_W^2 W_\mu^+ W^{-\mu}
- m_Z^2 Z_\mu Z^\mu \nonumber \\
+ (2 m_h^2 h^2 - \partial_\mu h \partial^\mu h) + \cdots
\eea
Inside the supernovae, the relevant matter fields that can be found are the nucleons, electrons-positrons, and plasmons. The interaction of these particles with the braneworld radion ($\phi$) is given in Eq. \eqref{interaction}. 

As discussed above, the process of our interest for the supernova SN1987A cooling is the neutrino pair production via the  $s$-channel exchange of a light stabilized radion  $\phi$ produced in a plasmon-plasmon collision:  
\beq
\gamma_P (k_1) + \gamma_P (k_2)  \stackrel{\phi}{\longrightarrow}   \nu_\ell (p_3) + \overline{\nu}_\ell(p_4),\;\;\;\ell = e, \mu, \tau.
\eeq 
where $k_1,~k_2$ are the incoming momenta and $p_3,~p_4$ are the outgoing momenta.

 Plasmons, which are the quanta of the electromagnetic fields in plasma, are massive and radion couples to plasmon which is proportional to the plasmon mass $m_A$.
The plasmon-plasmon-radion $\gamma_P-\gamma_P-\phi$ interaction vertex is given by \cite{GW}
\beq
- \frac{2 i m_A^2}{\langle \phi \rangle} \eta^{\mu \nu}
\eeq
where, $m_A$ is the plasmon mass and $\langle \phi \rangle$, the radion vev.  
The radion-fermion-fermion $f(p_3) - \overline{f}(p_4) - \phi(P)$ interaction vertex is given by \cite{GW}
\beq
- \frac{3 i}{2\vphi} \left[\slashed{p}_3 - \slashed{p}_4 - \frac{8}{3} m_f\right]
\eeq 
where $m_f$ is the mass of the fermion.  \\

\noindent Now for a generic $2$ body scattering $1+2 \to 3+4$, the scattering cross section is given by
\bea \label{Eq:sigma}
\sigma= \frac{1}{Flux} \int \prod_f \frac{d^3 p_f}{(2 \pi)^3 2 E_f} (2 \pi)^4 \delta^4 \left(p_3 + p_4 - k_1 - k_2\right) \overline{|{\cal{M}}|^2}
\eea 
where $Flux= 4 E_1 E_2 |v_{rel}| = 4 k \sqrt{S}$. Here $E_1$, $E_2$ are the energies of the two incoming plasmons, $v_{rel}$ is the relative velocity between them, and the $3$ momentum magnitude $|\bf{k}|$ is defined above.  \\
The energy loss per unit mass (in $erg~g^{-1}~s^{-1}$ unit) is defined as in a generic $2\to2$ body scattering process~\cite{Raffelt},
$$ \dot{\varepsilon} = \frac{Q_{a+b \to c+d}}{\rho_{SN}} = \frac{\langle n_a n_b \sigma_{a+b \to c+d} v_{rel} E_{cm}\rangle}{\rho_{SN}}$$
where $E_{cm} = E_a + E_b$, with $E_a, E_b$ are the energy of the two colliding particles $a$ and $b$, $n_a,~n_b$ are the number density, and $\rho_{SN}$ is the mass density.
\paragraph{$q$-deformed Statistics --} The plasmon number density in $q$-deformed statistics is given by \cite{Das} 
\begin{eqnarray}
n_{\gamma_{P}} = \int_{\omega_0}^{\infty} d\omega_1 \frac{\omega_1 (\omega_1^2 - \omega_0^2)^{1/2}}{\left[1 + (q - 1) b \omega\right]^\tau - 1}
\end{eqnarray}
The energy-loss rate due to plasmon-plasmon annihilation to neutrino pairs in $q$-deformed statistics can be written as
\begin{eqnarray}
\dot{\varepsilon} = \frac{Q}{\rho_{SN}} = \frac{1}{\rho_{SN}} \frac{1}{\pi^4} \int_{\omega_0}^{\infty} d\omega_1 \frac{\omega_1 \left(\omega_1^2 - \omega_0^2\right)^{\frac{1}{2}}}{\left[1 + (q - 1) b \omega_1\right]^\tau - 1}&&\int_{\omega_0}^{\infty} d\omega_2 \frac{\omega_2 \left(\omega_2^2 - \omega_0^2\right)^{\frac{1}{2}}}{\left[1 + (q - 1) b \omega_2\right]^\tau - 1  } \nonumber\\
&&\times\frac{s(\omega_1 + \omega_2)}{2 \omega_1 \omega_2}  ~~\sigma_{\gamma_P \gamma_P \to \nu_l \overline{\nu}_l}
\end{eqnarray}
where $E_1 = \omega_1$, $E_2 = \omega_2$, the plasmon mass $m_A = \omega_0$,  $\tau = \frac{1}{q-1}$ and $b = \frac{\beta_0}{4 - 3 q}$ (as defined earlier). The cross-section $\sigma_{\gamma_P \gamma_P \to \nu_l \overline{\nu}_l}$ is given by Eq.\ref{Eq:sigma} with the squared amplitude is given as 
\bea 
\overline{|M|^{2}}  = \frac{1}{3^2} \sum_{spins} |M|^{2} =  \frac{16 m_{A}^4 m_{\nu_l}^2}{{9 \langle \phi \rangle}^4} \frac{(s-4 m_{\nu_l}^2)}{\left[(s-m_\phi^2)^2 + m_\phi^2 \Gamma_\phi^2\right]} \left[1 + \frac{1}{2}\left(1 - \frac{s}{2 m_A^2}\right)^2\right] 
\label{Eq:amplsq}
\eea
Introducing the dimensionless variables $x_i = \om_i/T$($i=0,1,2$) and taking $m_A$ (the transverse plasmon mass) to be equal to $\omega_0$, we rewrite the energy loss rate at temperature $T$ as 
 \begin{equation}
 \dot{\varepsilon}  = \frac{1}{\rho_{SN}} \frac{T^7}{\pi^5} \int_{x_0}^{\infty} \int_{x_0}^{\infty} dx_1 dx_2 \frac{x_1 (x_1^2 - x_0^2)^{\frac{1}{2}}}{D_1}~ \frac{x_2 (x_2^2 - x_0^2)^{\frac{1}{2}}}{D_2 }~\frac{T^2 (x_1 + x_2)^3}{2 x_1 x_2} 
  \tilde{D}_1 \tilde{D}_2 ~  \overline{|M|^{2}}
 \end{equation}
 where $D_i = (1 + (q - 1) b x_i T)^\tau - 1,~\rm{i=1,2}$ and $\tilde{D}_j = 1 -\frac{1}{\left[1 + (q - 1) b x_j T\right]^\tau + 1}$. Here we set $\mu_{\nu_{l}} =0 $. The squared amplitude is given by Eq.~\ref{Eq:amplsq}. \\
 
\paragraph{Undeformed Statistics --} Writing the plasmon number density in the undeformed scenario,
\begin{eqnarray}
n_{\gamma_{P}} = \int \frac{ d^{3}k }{(2\pi)^{3}} g_{\gamma_{P}}~f(E) = \frac{1}{\pi^2} \int_{\omega_0}^{\infty} d\omega_1 \frac{\omega_1 (\omega_1^2 - \omega_0^2)^{\frac{1}{2}}}{e^{\omega_1/T} - 1}
\end{eqnarray}
the energy-loss rate due to plasmon-plasmon annihilation to neutrino pairs can be written as
\begin{equation}
\dot{\varepsilon} = \frac{Q}{\rho_{SN}} = \frac{1}{\rho_{SN}} \frac{1}{\pi^4} \int_{\omega_0}^{\infty} d\omega_1 \frac{\omega_1 (\omega_1^2 - \omega_0^2)^{\frac{1}{2}}}{e^{\omega_1/T} - 1}~ \int_{\omega_0}^{\infty} d\omega_2 \frac{\omega_2 (\omega_2^2 - \omega_0^2)^{\frac{1}{2}}}{e^{\omega_2/T} - 1}~\frac{s(\omega_1 + \omega_2)}{2 \omega_1 \omega_2}  ~~\sigma_{\gamma_P \gamma_P \to \nu_l \overline{\nu}_l}
\end{equation}
As before, after introducing the dimensionless variables $x_i = \om_i/T$($i=0,1,2$), we rewrite the energy loss rate at temperature $T$  
 \begin{eqnarray}
 \dot{\varepsilon}  = \frac{1}{\rho_{SN}} \frac{T^7}{\pi^5} \int_{x_0}^{\infty} dx_1 \frac{x_1 \left(x_1^2 - x_0^2\right)^{\frac{1}{2}}}{e^{x_1} - 1}~ \int_{x_0}^{\infty} dx_2&&\frac{x_2 \left(x_2^2 - x_0^2\right)^{\frac{1}{2}}}{e^{x_2} - 1} \nonumber \\
  &&\times\frac{s(x_1 + x_2)}{2 x_1 x_2}\frac{e^{x_1 + x_2}}{\left[1 + e^{\frac{(x_1 + x_2)}{2}}\right]^2}
  \overline{|M|^{2}}
  \hphantom{00}
 \end{eqnarray}
 where the squared amplitude  $ \overline{|M|^{2}}$ is defined above.

\section{Numerical Analysis} \label{sec:IV}
We now see how a light radion produced in a plasmon-plasmon collision emits neutrinos and takes part in the supernova SN1987A cooling. The process of our concern is $\gamma_P + \gamma_P \stackrel{\phi}{\longrightarrow}   \nu_x + \overline{\nu}_x$ (with $x=e, \mu, \tau$). 
Now Raffelt's criteria on the supernova energy loss rate \cite{Raffelt} states that if the SN1987A cools off due to any new physics process, the emissivity rate for that process must be $\dot{\varepsilon} \le 10^{19}~{erg~g^{-1}~s^{-1}}~(\sim 7.288\times 10^{-27} ~\rm{GeV})$. Using this criterion, we can obtain  bounds on radion VEV $\langle \phi \rangle$ as a function of radion mass ($m_\phi$). As mentioned earlier, we will work within the framework of Tsallis statistics and study the $q$-dependence of the bound on $\langle \phi \rangle$ for a given radion mass $m_\phi$.   

\subsection{Bound on $\langle\phi\rangle$ from SN1987A cooling}
\begin{figure}[htb]
 \centering
   \includegraphics[width=8.4cm]{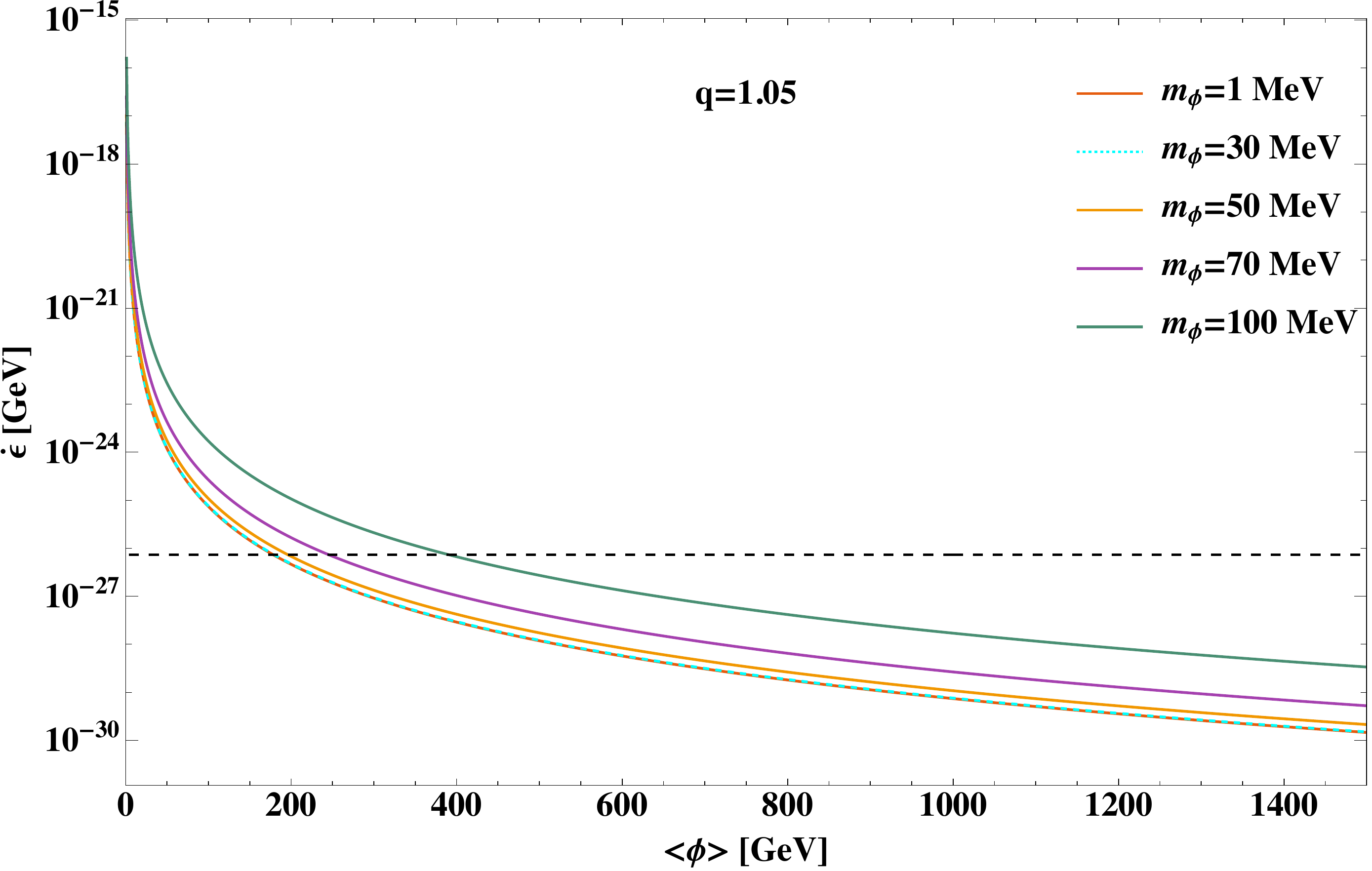}
   \includegraphics[width=8.4cm]{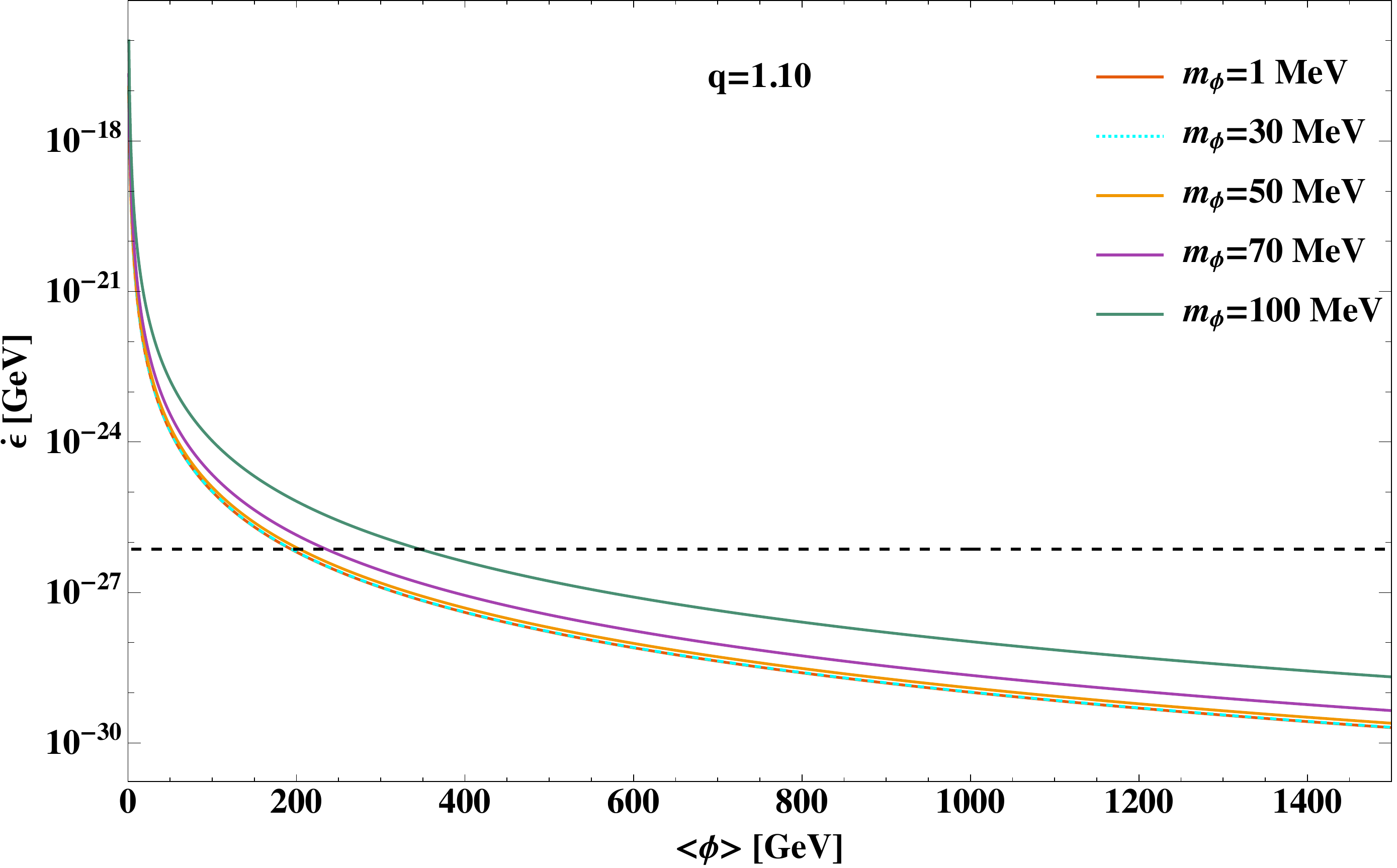}
      \caption{The SN1987A energy loss rate $\dot{\varepsilon} (\gamma_P\gamma_P  \stackrel{\phi}{\longrightarrow} \nu_\tau \overline{\nu}_\tau) $ is plotted against $\langle \phi \rangle$ (GeV) for different $m_\phi$ values. On the left, we set $q=1.05$, while on the right we set $q=1.10$. The horizontal line(dashed) in both plots corresponds to  the upper bound $\dot{\varepsilon} = 7.288\times 10^{-27} \rm{GeV}$. 
      }
 \label{Plot2}
  \end{figure}

~
In Fig.~\ref{Plot2}, we have plotted the energy-loss rate 
$ \dot{\varepsilon} (\gamma_P\gamma_P \stackrel{\phi}{\longrightarrow}  \nu_x \overline{\nu}_x)$ (with $x = \tau$) as a function of the radion VEV $\vphi$ for different $m_\phi$ and $q$ values\footnote{Since the radion coupling to on-shell fermions is proportional to its mass and the fact that $m_{\nu_e}, m_{\nu_\mu} \ll m_{\nu_\tau}$, the energy loss rate due to the $\nu_e$ and $\nu_\mu$ channel will be much less than that the $\nu_\tau$ channel and hence the former two channels (which comprises $\nu_e, ~\nu_\mu$ ) will not provide useful bound and the subsequent plots are not shown. However, the bound on $\vphi$ that follows from these two channels is shown in Table 2. }. On the left panel, we set $q=1.05$, while on the right, we set $q=1.10$.  
On each plot, the horizontal line  corresponds to the upper bound on the energy loss rate $\dot{\varepsilon} = 10^{19}~{erg~g^{-1}~s^{-1}}$(Raffelt's criterion). The region below the horizontal line is allowed. 
For $q=1.05(1.10)$ (in the \emph{left} (\emph{right}) plot in Fig.~\ref{Plot2}), we find the lower bound on $\vphi = 197\;(200)$ GeV  for $m_\phi = 50$ MeV, and   $\vphi = 407\;(408)$ GeV for $m_\phi = 100$ MeV, respectively. \\
Below in Fig.~\ref{Plot3} , we have presented a contour plot in the $q-\vphi$ plane  corresponding to the energy loss rate $\dot{\varepsilon} = 10^{19}~{erg~g^{-1}~s^{-1}}$ for the channels $(\gamma_P\gamma_P \stackrel{\phi}{\longrightarrow}  \nu_x \overline{\nu}_x)$ ($x=e, \mu, \tau$) and for different radion mass.  
\begin{figure}[!t]
 \centering
    \includegraphics[width=0.495\linewidth]{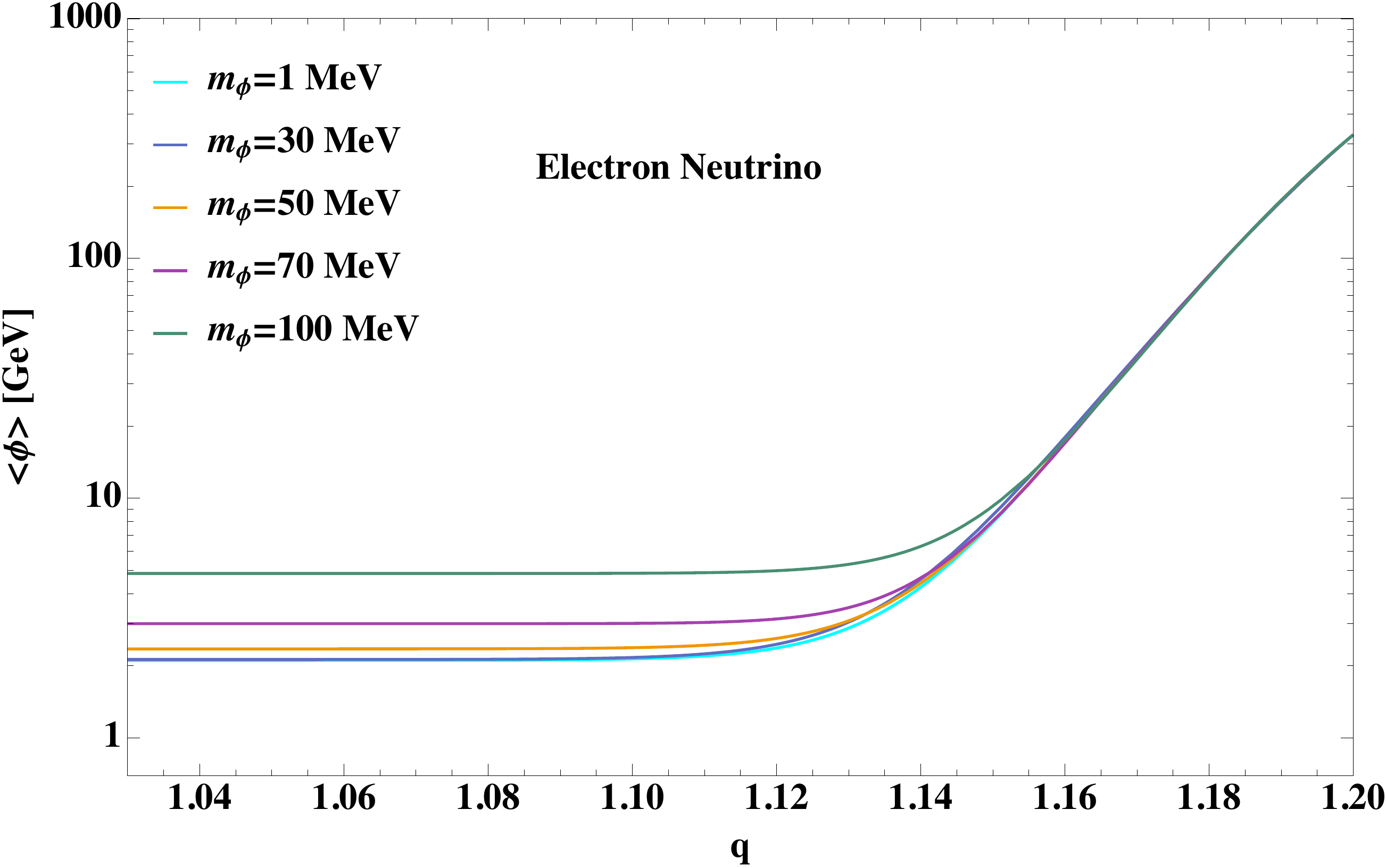}
    \includegraphics[width=0.495\linewidth]{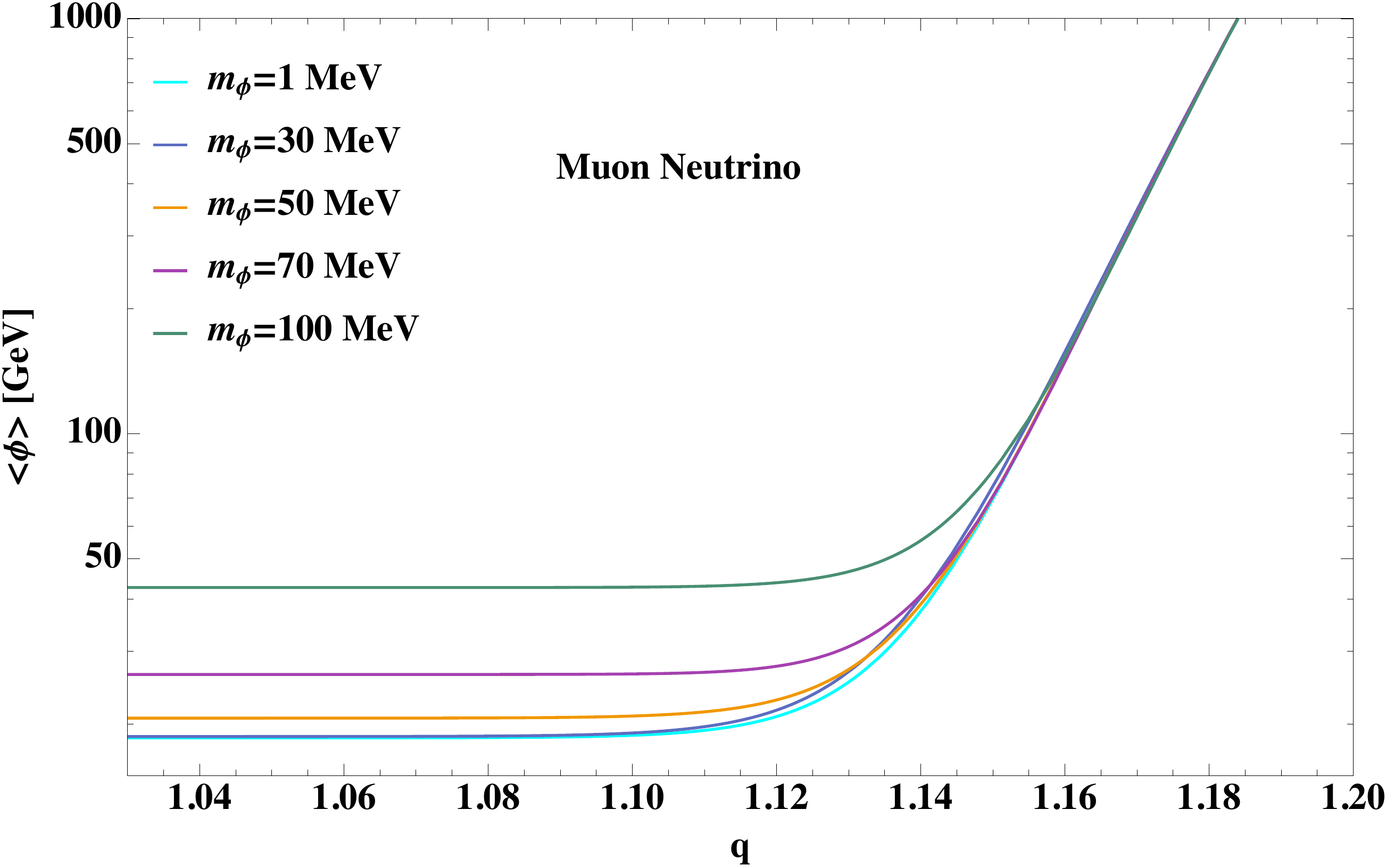}
    \includegraphics[width=0.495\linewidth]{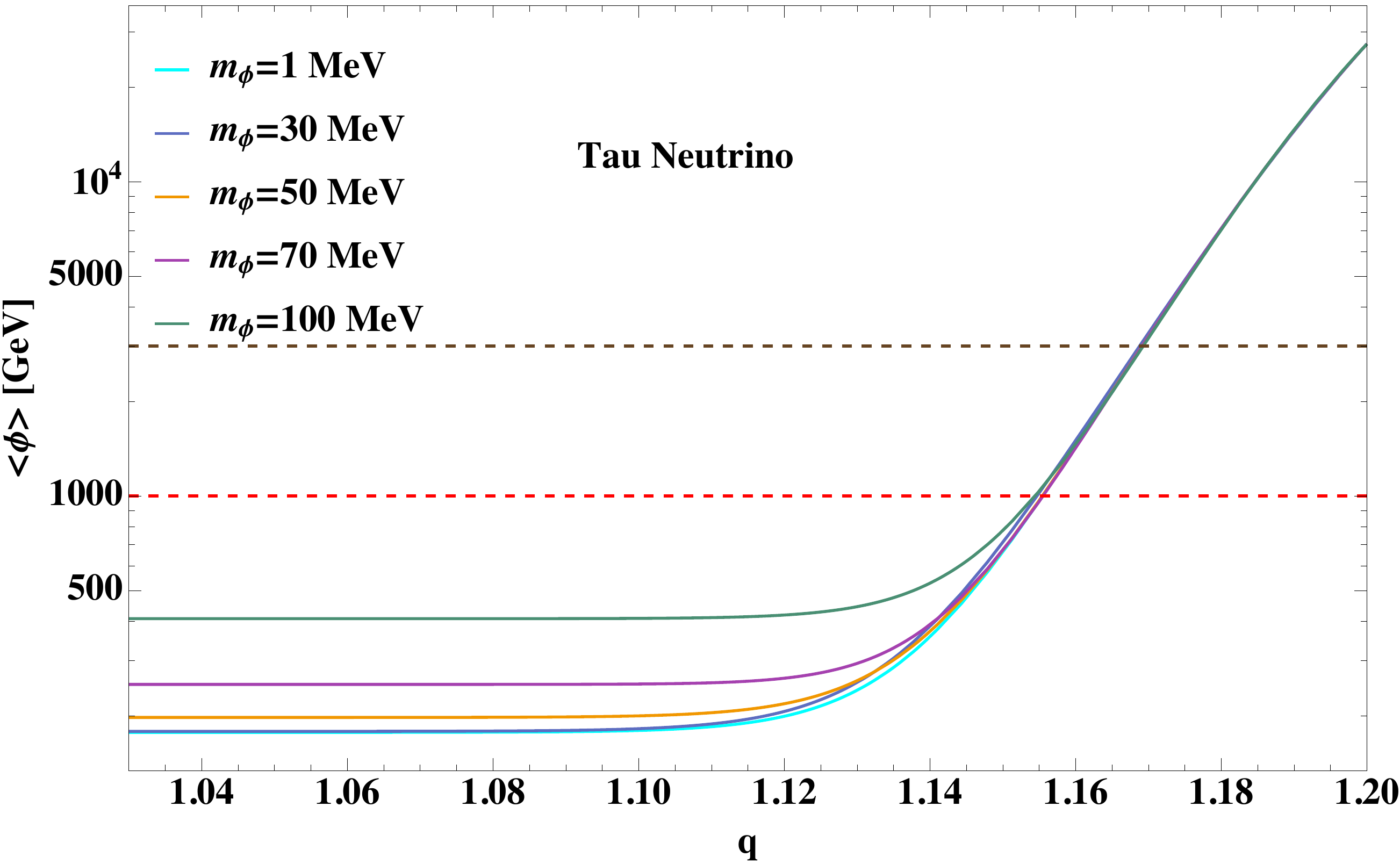}
  \caption{Contour plot in the $q - \langle \phi \rangle$ plane corresponding to different radion mass $m_\phi$. The bound on $\vphi$ follows from the upper bound on the energy loss rate i.e. $\dot{\varepsilon} (\gamma_P\gamma_P  \stackrel{\phi}{\longrightarrow}   \nu_x \overline{\nu}_x)=7.288 \times 10^{-27} \rm{GeV}$. Here $x = e, \mu, \tau$, respectively. The horizontal lines correspond to $\vphi = 1,~3$ TeV, respectively. }
 \label{Plot3}
  \end{figure}
From the figures, we see that for a given $m_\phi$, as $q$ increases, the lower bound on  $\vphi$first stays at a constant value and then starts increasing with $q$. In particular, we note that after $q=1.10$ the bound on $\vphi$ increases more rapidly.  The bound $\vphi$ that follows from $\nu_e$ and $\nu_\mu$ channels are found to be rather weaker in comparison to the one that follows from $\nu_\tau$ channel.
We have summarized the $q$-dependence of the lower bound on $\vphi$ for different $m_\phi$ values in Table~\ref{tab:table-1}.
\begin{table}[htb]
    \centering
    \resizebox{\textwidth}{!}{%
    \begin{tabular}{|c|ccc|ccc|ccc|ccc|ccc|}\hline
    & \multicolumn{15}{|c|}{$\mathbf{\langle \phi \rangle}$ [GeV]} \\ \cline{2-16}
    $M_\phi$ & \multicolumn{3}{c|}{$q=$1.05} & \multicolumn{3}{c|}{$q=$1.10} & \multicolumn{3}{c|}{$q=$1.15} & \multicolumn{3}{c|}{$q=$1.17} & \multicolumn{3}{c|}{$q=$1.20} \\ \cline{2-16}
    [MeV] & $e$ & $\mu$ & $\tau$ & $e$ & $\mu$ & $\tau$ & $e$ & $\mu$ & $\tau$ & $e$ & $\mu$ & $\tau$ & $e$ & $\mu$ & $\tau$ \\ \hline\hline
    1  &2.10 &18.51 & 176.80&2.13 &18.77 & 179.32&7.89 &69.43 &662.97 &38.46 &338.12 &3228.58 &327.21 &2876.34 &27465.10 \\
    30.00  &2.12 &18.65 &178.16 &2.16 &19.02 &181.65 &8.47 &74.47 &711.17 &39.31 &345.59 &3299.97 &327.58 &2879.60 &27496.20 \\
    50.00  &2.35 &20.65 &197.26 &2.37 &20.90 &199.56 &7.99 &70.28 &671.08 &38.37 &337.33 &3221.00 &327.18 &2876.13 &27462.60 \\
    70.00  &2.99 &26.32 &251.32 &3.00 &26.39 &252.04 &8.00 &70.34 &671.65 &38.35 &337.14 &3219.25 &327.17 &2876.03 &27461.50 \\
    100.00  &4.84 &42.61 &406.91 &4.85 &42.70 &407.77 &9.24 &81.28 &776.13 &37.86 &332.81 &3177.77 &327.01 &2874.59 &27447.10 \\ \hline
    \end{tabular}
    }
    \caption{The lower bound on the radion vev $\vphi$ that follows from the energy loss rate i.e. $\dot{\varepsilon} (\gamma_P\gamma_P  \stackrel{\phi}{\longrightarrow}   \nu_x \overline{\nu}_x)=7.288 \times 10^{-27} \rm{GeV}$ with $x = e, \mu, \tau$, respectively, are shown corresponding to different radion mass($m_\phi$) and different values of the deformation parameter $q$. }
    \label{tab:table-1}
\end{table}
From the table, we see that, for $m_\phi = 30$ MeV, for the channel $ \gamma_P\gamma_P  \stackrel{\phi}{\longrightarrow}   \nu_\tau \overline{\nu}_\tau$, as the deformation parameter $q$ varies from $q=1.05$ to $q=1.2$, the lower bound on $\vphi$ changes from $197$ GeV to $27496$ GeV, while that for electron and muon neutrino production channels, it varies from $2(18)$ GeV to $327(2879)$ GeV, respectively. One can also note that for the $\nu_e$ and $\nu_\mu$ channels, the lower bound on $\vphi$ that follows from the energy-loss rate is almost independent of the radion mass, however not so for $\nu_\tau$ channel.
Also, for a given $q$, as the radion mass $m_\phi$ increases, the lower bound on $\vphi$ also increases. As an example, for $q=1.15$, corresponding to the $\nu_\tau$ channel, as $m_\phi$ ranges from $1$ MeV to $100$ MeV, the lower bound on $\vphi$ increases from $662$ GeV to $776$ GeV, such dependence for higher $q$ values, say $q=1.17$ and $1.2$ are slightly different. \\
Finally, in the limit, $q \to 1$, the Tsallis statistical distribution formula takes the conventional Bose-Einstein or Fermi-Dirac statistical distribution formula i.e. 
\bea
f_i(\beta,E_i) = \frac{1}{\left[1 + (q-1)b E_i \right]^{\frac{1}{q-1}} \pm 1} ~\stackrel{q \to 1}{\longrightarrow} \frac{1}{e^{b E_i} \pm 1} \left(= \frac{1}{e^{\beta_0 E_i} \pm 1}\right)
\eea
where $e^{b E_i} = e^{\beta_0 E_i} $ with $b = \frac{\beta_0}{4 - 3 q} = \beta_0$ for $q \to 1$. Here $\beta_0$ is the inverse equilibrium temperature $T_0$ of the supernovae core. 
\begin{figure}[htb]
 \centering
  \includegraphics[width=0.6\linewidth]{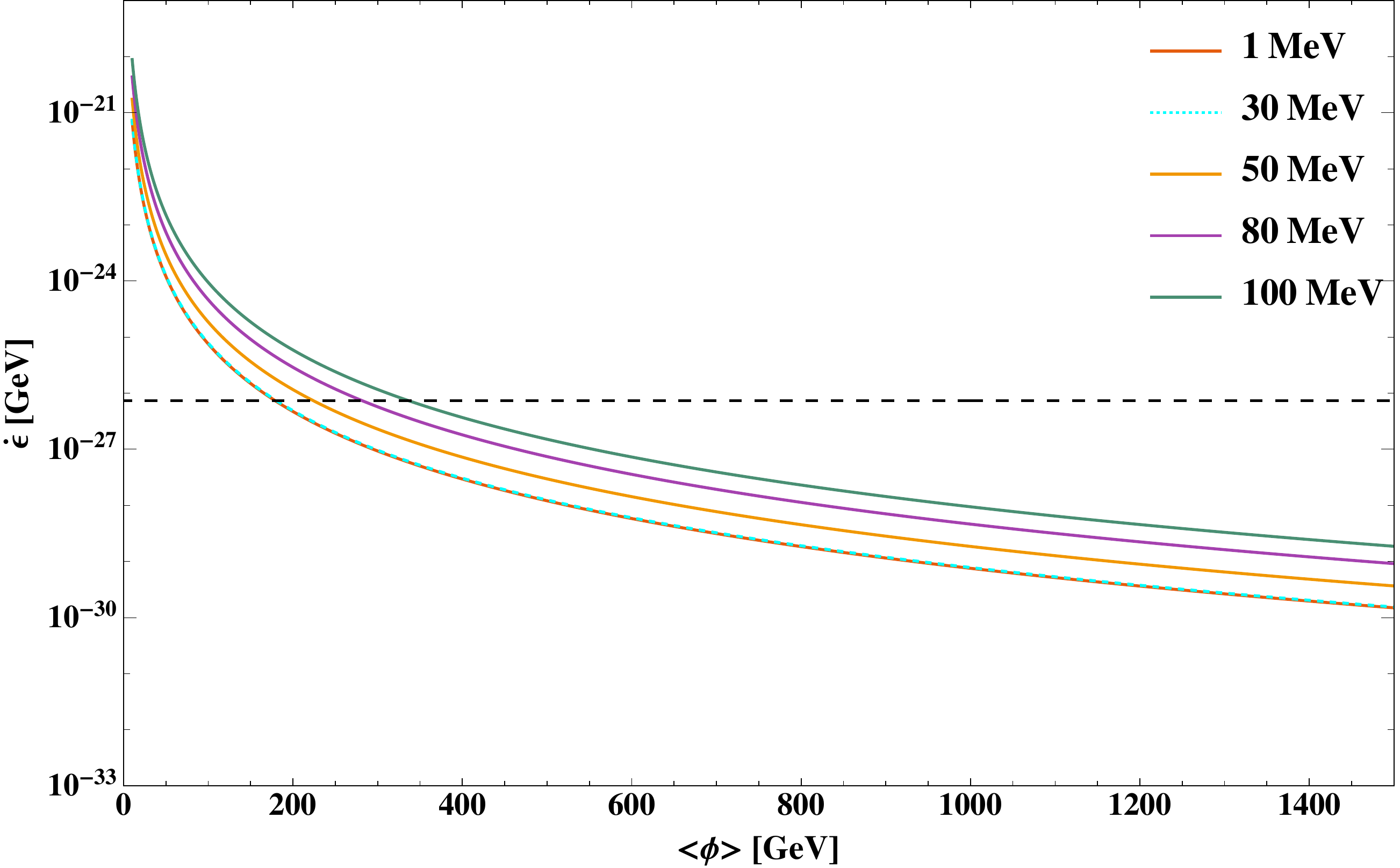}
  \caption{{The SN1987A energy loss rate due to the process $\gamma_P\gamma_P  \stackrel{\phi}{\longrightarrow}   \nu_\tau\overline{\nu}_\tau $ is plotted against the radion VEV, $\langle \phi \rangle$ for different $m_\phi$ values in undeformed framework (\textit{i.e.}, $q=1$).}}
 \label{Plot4}
  \end{figure}
In Fig. \ref{Plot4}, we have shown the SN1987A energy loss rate $\dot{\varepsilon} (\gamma_P + \gamma_P  \stackrel{\phi}{\longrightarrow}   \nu_\tau + \overline{\nu_\tau}) $ against $\vphi$ for different $m_\phi$ in undeformed scenario ($q=1$). The horizontal line corresponds to the upper bound of the energy loss rate i.e. $\dot{\varepsilon} (\gamma_P\gamma_P  \stackrel{\phi}{\longrightarrow}   \nu_\tau \overline{\nu}_\tau) = 7.288\times 10^{-27} \rm{GeV}$ and this gives a  lower bound on the radion VEV $\vphi = 178\, (407)~\rm{GeV}$ for $m_\phi = 30\, (100)~\rm{MeV}$ with $q=1$.  So, the lower bound on $\vphi$ increases with the increase in radion mass $m_\phi$. 
In Table 3, we have shown the lower bound on $\vphi$ corresponding to the supernova energy-loss rate $\dot{\varepsilon}(\gamma_P\gamma_P  \stackrel{\phi}{\longrightarrow}  \nu_{\tau} \overline{\nu}_{\tau}) \leq 7.288\times 10^{-27} \rm{GeV}$ for $q=1$ for different $m_\phi$ values.  
\begin{table}[htb]
    \centering
    \small
    \begin{tabular}{|c|ccc|}\hline
     
    $M_\phi$ & \multicolumn{3}{|c|}{$\mathbf{\langle \phi \rangle}_{q=1}$ [GeV]} \\ \cline{2-4}
    [MeV] & $e$ & $\mu$ & $\tau$ \\ \hline\hline
    1  &2.10 &18.52&176.79 \\
    30.00  &2.12 &18.66&178.14  \\
    50.00  &2.35 &20.66&197.25 \\
    70.00  &2.99 &26.32&251.32 \\
    100.00  &4.84 &42.62&406.91 \\ \hline
    \end{tabular}
    \caption{The lower bound on the radion vev $\vphi$ that follows from the energy loss rate i.e. $\dot{\varepsilon} (\gamma_P\gamma_P  \stackrel{\phi}{\longrightarrow}   \nu_x \overline{\nu}_x)=7.288 \times 10^{-27} \rm{GeV}$ with $x = e, \mu, \tau$, respectively, are shown corresponding to different radion mass($m_\phi$) and different values of the deformation parameter $q$. }
    \label{tab:table-3}
\end{table}
From Table 3, we see that as $m_\phi$ increases from $1$ MeV to $100$ MeV, the bound on $\vphi$ increases from $177$ GeV to $407$ GeV, respectively.

 \section{Conclusion} \label{sec:V}
We have studied the impact of a light-stabilized Randall-Sundrum radion in the supernova SN1987A cooling. The radion produced due to plasmon-plasmon collisions in the outer crust of the supernova core can decay into a pair of neutrinos which take away energy released in the supernova SN1987A explosion. Assuming that the energy loss rate $\dot{\varepsilon}(\gamma_P\gamma_P  \stackrel{\phi}{\longrightarrow}  \nu_{\tau} \overline{\nu}_{\tau}) \leq 7.288\times 10^{-27} \rm{GeV}$, we obtain a lower bound on the radion vev $\vphi$ which depends on the radion mass $m_\phi$ and the deformation parameter $q$.  We find the lower bound on $\vphi  \sim 197$ GeV, $200$ GeV, $671$ GeV, $3221$ GeV and $27462$ GeV for $m_{\phi} = 50$ MeV corresponding to $q=1.05,\; 1.10,\; 1.15,\; 1.17$ and $1.2$, respectively. In the $q=1$ scenario (undeformed), we find the lower bound on $\vphi = 197 (407)$ GeV corresponding to $m_{\phi} = 50 (100)$ MeV. 

\section*{Acknowledgment}
PKD would like to thank Uma Mohanta who introduced the braneworld radion to him.  M K Sharma and S Kundu would like to acknowledge the fellowship support of BITS Pilani K K Birla Goa Campus.


 \bibliographystyle{unsrt} 
 \bibliography{Reference}

\end{document}